\documentstyle[emulateapj,psfig]{article}
\def\gs{\mathrel{\raise0.35ex\hbox{$\scriptstyle >$}\kern-0.6em 
\lower0.40ex\hbox{{$\scriptstyle \sim$}}}}
\def\ls{\mathrel{\raise0.35ex\hbox{$\scriptstyle <$}\kern-0.6em 
\lower0.40ex\hbox{{$\scriptstyle \sim$}}}}

\submitted{ }
 
\lefthead{IVISON ET AL.}
\righthead{DUST, GAS AND THE EVOLUTIONARY STATUS OF 8C\,1435+635}

\begin{document}
\title{Dust, gas and the evolutionary status of the 
       radio galaxy, 8C\,1435+635, at {\scriptsize z} = 4.25}

\author{R.\ J.\ Ivison\altaffilmark{1},
        J.\ S.\ Dunlop,
        D.\ H.\ Hughes and
        E.\ N.\ Archibald}
\affil{Institute for Astronomy, Department of Physics \& Astronomy, 
       University of Edinburgh, Blackford Hill, Edinburgh EH9 3HJ}

\author{J.\ A.\ Stevens,
        W.\ S.\ Holland and
        E.\ I.\ Robson}
\affil{Joint Astronomy Centre, 660 North A'oh\=ok\=u Place,
       University Park, Hilo, HI 96720}

\author{S.\ A.\ Eales}
\affil{Department of Physics, University of Wales, College of
       Cardiff, PO Box 913, Cardiff CF4 3TH}

\author{S.\ Rawlings}
\affil{Astrophysics, Nuclear Physics Laboratory, Oxford University,
       Keble Road, Oxford OX1 3RH}

\author{A.\ Dey}
\affil{KPNO/NOAO, 950 N.\ Cherry Avenue, PO Box 26732, Tucson,
       AZ 85726}

\and

\author{W.\ K.\ Gear}
\affil{Royal Observatory, Blackford Hill, Edinburgh EH9 3HJ}

\altaffiltext{1}{Supported by a PPARC Advanced Fellowship.}

\begin{abstract}
We present the results of new rest-frame far-IR observations of the $z
= 4.25$ radio galaxy, 8C\,1435+635, which not only confirm that it
contains an enormous quantity of dust (as first inferred from its
mm-wave detection by Ivison 1995) but also allow the first meaningful
constraints to be placed on the mass of this dust and associated gas.
The new measurements consist of: (i) clear detections of sub-mm
continuum emission at $\lambda_{\rm obs} = 450$ and $850$\,$\mu$m
obtained with the new sub-mm bolometer array, SCUBA, on the James
Clerk Maxwell Telescope (JCMT), (ii) continuum upper limits at
$\lambda_{\rm obs} = 350$, $750$ and $175$\,$\mu$m obtained with SCUBA
and the PHT far-IR camera aboard the {\em Infrared Space Observatory}
({\em ISO}), and (iii) a sensitive upper limit on the CO($4-3$) line
flux obtained with the IRAM 30-m MRT.  The resulting rest-frame
33---238-$\mu$m continuum coverage allows us to deduce that $2 \times
10^8$\,M$_{\odot}$ of dust at a temperature of $40 \pm 5$\,{\sc k} is
responsible for the observed mm/sub-mm emission. Using our CO upper
limit, which constrains $M_{{\rm H}_2}/M_{\rm d}$ to $<950$, we go on
to calculate robust limits on the total gas reserves (H$_2$ + H\,{\sc
i}) which are thereby constrained to between $4 \times 10^{10}$ and
$1.2 \times 10^{12}$\,M$_{\odot}$. The sub-mm properties of
8C\,1435$+$635 are thus strikingly similar to those of the $z = 3.80$
radio galaxy, 4C\,41.17, the only other high-redshift galaxy detected
to date at sub-mm wavelengths whose properties appear {\em not} to be
exaggerated by gravitational lensing (Dunlop et al.\ 1994; Hughes,
Dunlop \& Rawlings 1997).  The inferred gas masses of both objects are
sufficiently large to suggest that the formative starbursts of massive
elliptical galaxies are still in progress at $z \simeq
4$. Observations of complete samples of radio galaxies spanning a
range of redshifts and radio luminosities will be required to
determine if the spectacular far-IR properties of 8C\,1435$+$635 and
4C\,41.17 are primarily due to their extreme redshifts or their
extreme radio luminosities.
\end{abstract}


\section{Introduction}

Systematic redshift surveys out to $z \simeq 1$, and the discovery of
Lyman-limit galaxies at $z \simeq 3$, have allowed considerable
progress to be made in understanding the star-formation history of
some bright, present-day galaxies, and suggest that global
star-formation activity in our Universe peaked at around $z \simeq 2$
(e.g.\ Lilly et al.\ 1995; Pei \& Fall 1995; Madau et al.\ 1996;
though see Smail, Ivison \& Blain 1997).  However, these same optical
surveys appear to confirm that the properties of massive elliptical
galaxies are little changed by $z \simeq 1$, consistent with the
picture that most of the stars in massive ellipticals were formed in a
relatively short-lived, intense starburst at high redshift (e.g.\
Dunlop et al.\ 1996).  If, as has been suggested by a number of
authors (e.g.\ Zepf \& Silk 1996), this initial starburst is biased
towards the formation of significant quantities of {\em high-mass}
stars (and hence dust) then a massive elliptical in the throes of
formation would be expected to emit copious quantities of far-IR
radiation. When viewed at high redshift, such an object would
therefore be expected to be a strong sub-mm source. Consequently it
has long been anticipated that the formation and evolution of
elliptical galaxies will be one of the key cosmological issues which
can be best addressed through the introduction of deep sub-mm imaging.

Despite the success of the optical surveys described above in sampling
the Universe out to $z \simeq 3$, and the exciting potential of deep
sub-mm surveys to probe the high-redshift Universe, the study of radio
galaxies remains arguably the best method by which to trace the
cosmological evolution of massive elliptical galaxies. This is because
selection on the basis of strong {\it extended} radio emission
guarantees: (i) that the host galaxy is a giant elliptical or the
progenitor thereof (Matthews, Morgan \& Schmidt 1964; Lilly \& Longair
1984), a key advantage over optical/IR surveys where one is forced
to resort to circumstantial evidence, such as comoving number
densities, to help make an educated guess as to the most appropriate
low-redshift counterpart for a given high-redshift source; (ii)
highly efficient selection of high-redshift galaxies, the properties
of which should {\it not} be significantly biased by gravitational
lensing (in contrast, for example, to optically selected quasars which
are selected primarily on the basis of their strong, {\em compact}
emission).
 
These considerations suggest that steep-spectrum radio galaxies are
currently the targets of choice to trace the cosmological evolution of
dust and gas (and hence star-formation activity) in massive elliptical
galaxies. Furthermore, such reasoning has already received a
considerable observational boost as a result of the successful
mm/sub-mm detections of first 4C\,41.17 (Dunlop et al.\ 1994; Chini \&
Kr\"ugel 1994) and then 8C\,1435$+$635 (Ivison 1995).

The most useful physical parameter which can be extracted from such
observations is the {\em mass} of dust, and hence the amount of gas
which remains to be converted into stars at the epoch of observation.
However, as discussed in detail by Hughes et al.\ (1997), to obtain a
reliable estimate of the dust and gas masses in high-redshift objects
requires that the uncertainty in dust temperature be minimized through
obtaining mm---far-IR measurements which extend to significantly
shorter wavelengths than the Rayleigh-Jeans tail, preferably
straddling the rest-frame far-IR emission peak. Furthermore, it is
helpful (through observations of CO lines) to confirm that gas/dust
ratios in such high-redshift sources are at least consistent with
those seen in low-redshift galaxies, before daring to extrapolate from
the dust mass to the total mass of gas available for future star
formation.
 
In this paper we address both these issues through a concerted
programme of new mm/sub-mm/far-IR observations of the $z = 4.25$ radio
galaxy, 8C\,1435$+$635. Such a programme has only been made possible
through the advent of new, highly sensitive facilities, in particular
the newly-commissioned sub-mm bolometer array, SCUBA, on the
JCMT\footnote{The JCMT is operated by the Observatories on behalf of
the UK Particle Physics and Astronomy Research Council, the
Netherlands Organization for Scientific Research and the Canadian
National Research Council.}, and the PHT far-IR camera aboard {\em
ISO}. Together these facilities offer the opportunity of moving
mm/far-IR studies of high-redshift objects from the pioneering world
of bare detections to the reliable extraction of meaningful physical
parameters.

The layout of this paper is as follows. First we describe our existing
and new mm $\rightarrow$ far-IR measurements of 8C\,1435+635, which
prove beyond doubt that the far-IR emission from this galaxy is
produced by dust. Next, we use these data to derive new constraints on
the dust and inferred gas mass in this galaxy. We then discuss the
implications of our results for the evolutionary status of
8C\,1435$+$635, compare its properties with those of 4C\,41.17, and
conclude with a brief discussion of the implications of this work for
the evolution and formation of massive ellipticals in general.  Unless
otherwise stated we assume $q_0 = 0.5$ and $H_0 =
50$\,km\,s$^{-1}$\,Mpc$^{-1}$ throughout; the impact on our
conclusions of adopting a lower value for $q_0$ is specifically
addressed in \S4.

\section{Observations}

\subsection{8C\,1435+635; existing data}

The existence of a large mass of dust in 8C\,1435+635 was first
inferred as a result of its detection at $\lambda_{\rm obs} =
1250$\,$\mu$m ($S_{\rm 1250} = 2.57 \pm 0.42$\,mJy) by Ivison (1995).
At the time of this observation, 8C\,1435+635 (4C\,63.20) was the most
distant known radio galaxy, with a redshift $z = 4.25$ corresponding
to a look-back time of around 92 per cent of the age of the Universe.
Its prodigious radio luminosity ($P_{\rm 1.5\,GHz} = 5.4 \times
10^{27}$\,W\,Hz$^{-1}$\,sr$^{-1}$, Lacy et al.\ 1994) allied with a
surprisingly weak Lyman $\alpha$ line ($L_{{\rm Ly}\alpha} = 5.5
\times 10^{36}$\,W, Spinrad, Dey and Graham 1995), made 8C\,1435+635 a
natural target for observers in the sub-mm/mm regime where reservoirs
of dusty, molecular gas betraying the galaxy's initial, brief burst of
star formation might be seen.
 
A further attraction of this source as a potential target for
mm/sub-mm observations arises from the fact that, like 4C\,41.17
(Dunlop et al.\ 1994), 8C\,1435+635 is an ultra-steep spectrum (USS)
radio source, in which extrapolation of its radio spectrum leads to an
expected synchrotron contribution of less than 0.01\,mJy at
$\lambda_{\rm obs} \simeq 1$\,mm.  For this reason, the 1250-$\mu$m
emission observed by Ivison (1995) was difficult to explain as arising
from anything other than thermal radiation from dust. Nevertheless, it
is worth re-emphasizing that this one data-point was insufficient to
conclusively exclude some previously undetected high-frequency
synchrotron component from being responsible for the rest-frame far-IR
emission.

\subsection{Measurements with SCUBA on JCMT}

 \begin{table*}
 {\scriptsize
 \begin{center}
 Table 1

 \vspace{0.1cm}
 Photometry of 8c\,1435+635 using JCMT and {\em ISO}

 \vspace{0.3cm}
 \begin{tabular}{lccc}
 \hline\hline
 \noalign{\smallskip}
 {Detector} & {UT Date} & {Wavelength} & {Flux Density} \cr
 {}         & {(1997)}  & {/$\mu$m}    & {/mJy}         \cr
 \hline
 \noalign{\smallskip}
 PHT      &April 13    &175             &$3\sigma < 40.1$           \cr
 SCUBA    &May 12--13  &350             &$3\sigma < 87.0$           \cr
 SCUBA    &April--June &450             &$23.6 \pm 6.4$             \cr
 SCUBA    &May 12--13  &750             &$8.74 \pm 3.31$\tablenotemark{\dag}\cr
 SCUBA    &April--June &850             &$7.77 \pm 0.76$            \cr
 \noalign{\smallskip}
 \noalign{\hrule}
 \noalign{\smallskip}
 \end{tabular}

 $^{\dagger}$ $2.64$-$\sigma$ marginal detection; $3\sigma < 18.7$\,mJy.
 \end{center}
 }
 \vspace*{-0.8cm}
 \end{table*}

Data were obtained during the period 1997 April---June using the
0.1-{\sc k}, sub-mm common-user bolometer array (SCUBA --- Robson
et al.\ 1998). SCUBA has two arrays of bolometric detectors which are
operated at 0.1\,{\sc k} to achieve sky background-limited performance
on the telescope. The LW array operates at 750 and 850\,$\mu$m and has
37 pixels, each of which has a diffraction-limited beam of diameter
14$''$ at 850\,$\mu$m. The SW array has 91 pixels and operates at 350
and 450\,$\mu$m with beamwidths of 7.5$''$. Both arrays have a 2.3$'$
instantaneous field of view, and by means of a dichroic beamsplitter
one can observe at two sub-mm wavelengths simultaneously (e.g.\
350 + 750 or 450 + 850\,$\mu$m).
 
Photometry of point-like sources (i.e.\ those that are unresolved) is
performed using the central pixels of each array, which are aligned to
within an arcsecond of each other. Experience has shown that, for
good-to-moderate seeing conditions, the best photometric accuracy is
achieved by averaging the source signal over a slightly larger area
than the beam. This is achieved by `jiggling' the secondary mirror in
a filled-square, 9-pt pattern, with a 2$''$ offset between each
point. The integration time at each point in a jiggle is 1\,s, and so
the pattern takes 9\,s to complete. During the jiggle the secondary
mirror was chopped azimuthally by 60$''$ at 7\,Hz. After the first 9-s
jiggle, the telescope was nodded to the reference position
(subsequently every 18\,s).
 
Data at 450 and 850\,$\mu$m were obtained during June 12 {\sc ut};
measurements at 350, 450, 750 and 850\,$\mu$m, taken during the
commissioning of SCUBA in the period 1997 April---May, were also
used. In total, 280\,min were spent on source at 450 and 850\,$\mu$m
and 100\,min at 350 and 750\,$\mu$m.

Skydips were performed before, during and after the target
measurements. During June 12, the the atmospheric zenith opacities at
450 and 850\,$\mu$m were very stable: 0.66 and 0.16, respectively;
during the commissioning nights, the zenith opacity was in the range
1.65---1.96 at 350\,$\mu$m, 0.78---1.05 at 450\,$\mu$m, 0.51---0.70 at
750\,$\mu$m and 0.16---0.21 at 850\,$\mu$m. The airmass of
8C\,1435+635 was between 1.38 and 1.58. Telescope pointing accuracy
was checked regularly using 1308+326, Arp~220 and 1418+546 and the
largest pointing correction after two slews was 2.9$''$. All data were
calibrated against Mars.
 
\vspace*{-1cm}
\hbox{~}
\centerline{\psfig{file=f1.eps,angle=-90,width=4.8in}}
\vspace*{-0.7cm}
\noindent{\scriptsize
             Fig.~1.
             Central SCUBA bolometer signal at 850\,$\mu$m ({\em filled
             symbols}) and the average signal of the adjacent six bolometers
             ({\em open symbols}) over the course of forty 18-s
             integrations during 1997 May 02, clearly showing that the
             signal is dominated by spatially correlated sky emission. A
             similar trend was seen at 450\,$\mu$m.

}

\vspace*{0.5cm}
Data reduction consisted of taking the measurements from the central
bolometer, rejecting spikes, and averaging over 18\,s. Data from the
adjacent bolometers were treated in a similar manner. Figure~1 shows
that the scatter in values measured by the central bolometer is
repeated in the average of the adjacent detectors.  The signal
detected by all the bolometers is clearly dominated by spatially
correlated sky emission.
 
We experimented with several methods of removing the residual sky
background, using means and medians of $1$ to $n$ rings of `residual
sky background' pixels ($n=1,2,3$ at 850\,$\mu$m). These schemes all
gave consistent results. Removing the sky in this way reduced the
effective noise-equivalent flux density from 140 to
95\,mJy\,Hz$^{-1/2}$, which is in excellent agreement with our photon
noise-based models, suggesting that we have removed the effects of
excess sky noise entirely. Figure~2 shows data coadded over the course
of several nights, where the standard error integrates down with time
as $95 (t/{\rm s})^{-1/2}$\,mJy during a period of 17000\,s. The
measured flux densities are reported in Table~1.

\vspace*{-1cm}
\hbox{~}
\centerline{\psfig{file=f2.eps,angle=-90,width=4.8in}}
\vspace*{-0.7cm}
\noindent{\scriptsize
             Fig.~2.
             The evolution with time of the standard error in 850-$\mu$m 
             flux density as measured with SCUBA. The solid line shows
             the error falling as $95 (t/{\rm s})^{-1/2}$\,mJy and the
             data conform to this ideal.

}
\vspace*{0.5cm}

\subsection{Mapping at 175\,$\mu$m with {\em ISO}}

On 1997 April 13, during orbit 514 of {\em ISO}\footnote{ISO is an ESA
project with instruments funded by ESA Member States with the
participation of ISAS and NASA.}  (Kessler et al.\ 1996), we used the
far-IR camera, PHT (Lemke et al.\ 1996), to obtain an over-sampled map
centered on 8C\,1435+635. We used the C200 camera (a $2 \times 2$
stressed Ge:Ga array with 90$''$ pixels) and a broad-band filter
centered at 175\,$\mu$m to perform a $4 \times 3$ raster with 46$''$
grid spacings.  We integrated for 57\,s at each point in the grid and
the major axis of the $4.6' \times 5.4'$ map was at a position angle
of 108$^{\circ}$.  The C200 raster was preceded and followed by 32-s
calibration scans of an internal photometric standard.
 
The data were reduced using the PHT Interactive Analysis\footnote{PIA
is a joint development by the ESA Astrophysics Division and the
ISOPHOT Consortium.} package ({\sc pia} V6.0e) and an upper limit is
reported in Table~1. Corrections were made for detector non-linearity,
etc., and the data were extensively deglitched. Data quality was good,
with consistent calibration scans. Our maps were made using the best
available flat-fielding algorithms and calibration information, but
forthcoming {\sc pia} software may be able to improve upon our
reduction.

\subsection{CO measurements from the IRAM 30-m MRT}

8C\,1435+635 was observed during 1996 August 17--18 using the IRAM
30-m MRT. Two 512-MHz filterbanks were used as backends for the {\sc
3mm1} and {\sc 3mm2} receivers; the central frequency for each
receiver was 87.784\,GHz, giving coverage over the range, $4.237 < z <
4.267$ for the CO($4-3$) rotational transition (1700\,km\,s$^{-1}$ of
velocity coverage). The 25$''$ beam was nutated by 120$''$ in azimuth
at a rate of 0.5\,Hz, with the telescope position-switching by the
same distance every 30\,s to alternate the signal and reference
beams. Altogether, exclusive of overheads, 13.3\,hr were spent on
source. The atmospheric zenith opacity was normally around 0.1 and
$T_{\rm sys}$ was $\sim 250$ and $\sim 190$\,{\sc k} for {\sc 3mm1}
and {\sc 3mm2}.
 
The {\sc class} reduction package was used to calibrate the spectra on
the $T_{\rm MB}$ scale (where 1\,{\sc k} = 4.7\,Jy), then to coadd and
bin the data into 24-MHz channels (80\,km\,s$^{-1}$). We found that
the noise level in our coadded spectrum decreased steadily with
time. The final noise level was 0.3\,m{\sc k} (in line with
theoretical expectations). For a Gaussian line profile of width
300\,km\,s$^{-1}$, we derive a CO line luminosity of $3\sigma < 5
\times 10^{10}$\,{\sc k}\,km\,s$^{-1}$\,pc$^2$, or, for $M({\rm
H}_2)/L_{\rm CO}' = 4$\,M$_{\odot}$ ({\sc
k}\,km\,s$^{-1}$\,pc$^2)^{-1}$ (Evans et al.\ 1996), $M({\rm H}_2,
3\sigma) < 2 \times 10^{11}$\,M$_{\odot}$.

\section{Results}

The spectral energy distribution (SED) of 8C\,1435+635 is shown in
Figure~3. A search for CO($1-0$) using the Very Large Array (van Ojik et
al.\ 1997) extended the frequency coverage of the radio regime to
22\,GHz and shows that the radio spectral index continues to be as
steep and negative at 22\,GHz as it is between 4.9 and 15\,GHz.
 
\vspace*{0.5cm}
\hbox{~}
\centerline{\psfig{file=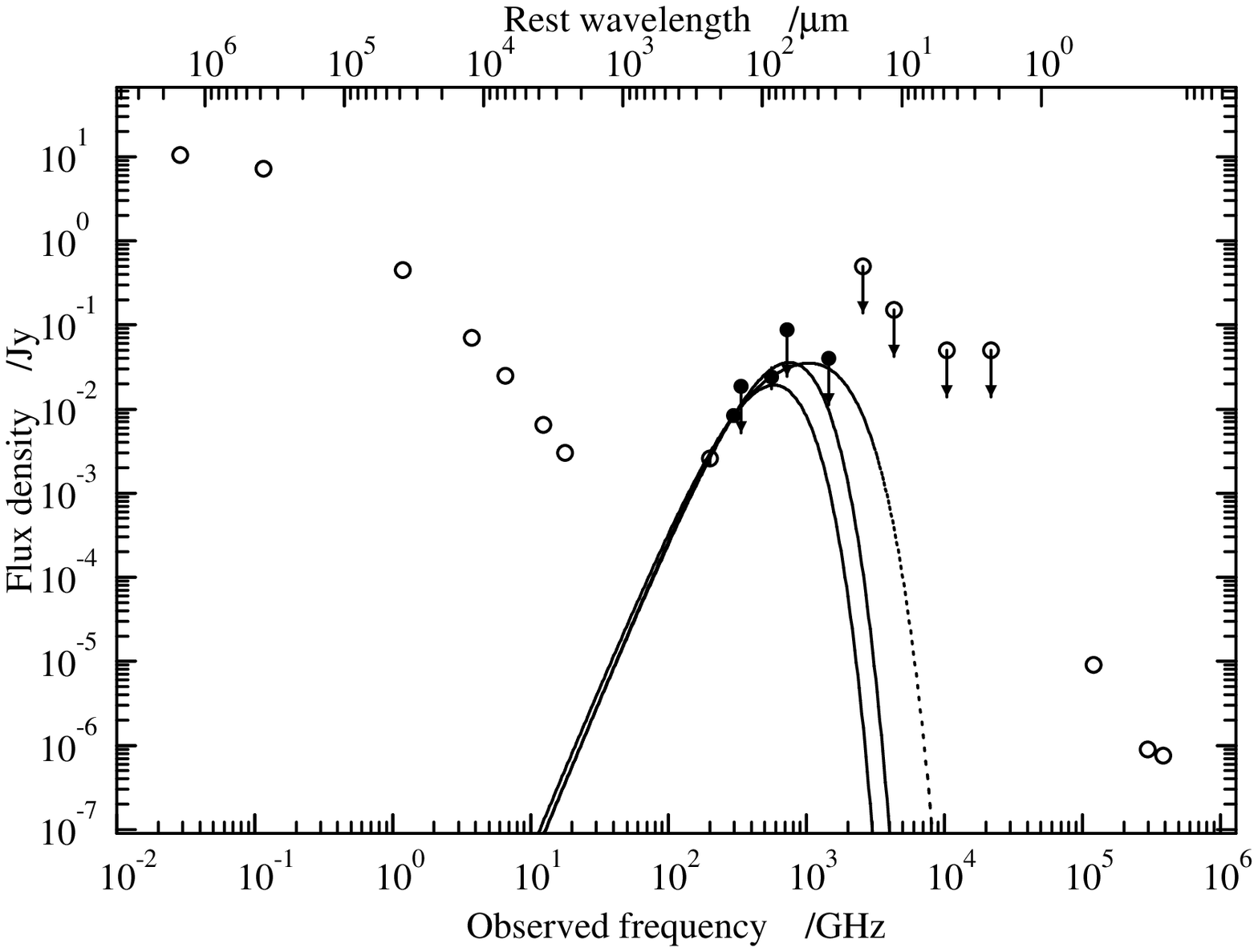,angle=0,width=2.8in}}
\vspace*{0.3cm}
\noindent{\scriptsize
             Fig.~3. SED of 8C1435+635. New data ({\em filled
             circles}); older data ({\em open circles}) from Hales et
             al.\ (1990), Rees (1990), Lacy et al.\ (1994), Spinrad et
             al.\ (1995), Ivison (1995), Carilli et al.\ (1997) and
             van Ojik et al.\ (1997). Two modified blackbodies ({\em
             solid lines}) represent dust at 35 and 45\,{\sc k}
             ($\beta=+2.0$) --- the most extreme temperatures
             commensurate with the 450-$\mu$m flux density for
             optically thin, isothermal dust. Another modified blackbody ({\em
             dotted line}) represents a 110-{\sc k}, isothermal dust
             component (also $\beta=+2.0$), still commensurate with
             the 175---1250-$\mu$m SED but which becomes optically
             thick at 200\,$\mu$m, demonstrating that the dust
             properties remain poorly constrained.

}
\vspace*{0.3cm}

Assuming that there is no flatter, high-frequency component, the
non-thermal contribution at the wavelengths of SCUBA's sub-mm
filters is less than 0.01\,mJy. This indicates that we can safely
interpret the mm $\rightarrow$ far-IR regime in isolation, and it is
immediately clear that this portion of the SED is dominated by
emission from dust.

 \begin{table*}
 {\scriptsize
 \begin{center}
 Table 2

 \vspace{0.1cm}
 Gas and dust parameters for high-redshift systems

 \vspace{0.3cm}
 \begin{tabular}{lcccc}
 \hline\hline
 \noalign{\smallskip}
 {Object}&{$L_{\rm CO}'$}&{$M_{{\rm H}_2}$}&{$L_{\rm FIR}$}&{$M_{\rm d}$} \cr
 {}&{/$10^9$\,K\,km\,s$^{-1}$\,pc$^2$}&{/$10^{9}$\,M$_{\odot}$}&
 {/$10^{12}$\,L$_{\odot}$}&{/$10^{6}$\,M$_{\odot}$} \cr
\hline
\noalign{\smallskip}
 FSC\,10214+4724&55       &25             &4.4          &105 \cr
 BR\,1202$-$0725&72       &330            &65           &1530\cr
 H\,1413+117    &24       &41             &11           &240 \cr
 53W002$^{\ddagger}$&42   &170            &12           &280 \cr
 4C\,41.17      &$<$65    &$<$310         &25           &580 \cr
 8C\,1435+635   &$<$50    &$<$200         &10           &215 \cr
\noalign{\smallskip}
\noalign{\hrule}
\noalign{\smallskip}
\end{tabular}

 $^{\ddagger}$ Marginal sub-mm detection (Hughes et al.\ 1997).

 \centerline{{\sc Note}.---Consistent use of: $q_0 = 0.5$; $H_0 =
 50$\,km\,s$^{-1}$\,Mpc$^{-1}$; Gaussian lines}
 \centerline{of width 300\,km\,s$^{-1}$; lensing factors --- H\,1413+117
 ($\times 11$); FSC\,10214+4724}
 \centerline{($\times 20$). See Ivison et al.\,(1996), Scoville et al.\,(1995,
 1997), Ohta et al.\,(1996),}
 \centerline{Barvainis et al.\ (1997).}
\end{center}
}
\vspace*{-0.8cm}
\end{table*}

\subsection{Dust properties}

Hughes et al.\ (1997) have shown that the uncertainties in calculating
the mass of dust responsible for the thermal, sub-mm emission in
high-redshift radio galaxies are: (i) our limited knowledge of the
rest-frame mass absorption coefficient, $\kappa_{\rm d}$, and its
dependence on frequency; (ii) the dust temperature, $T_{\rm d}$, which
has rarely been constrained to better than $\pm 20$\,{\sc k} for any
high-redshift system, even assuming optically thin dust, and, finally,
(iii) the elusive values of $H_0$ and $q_0$. Unfortunately, these
problems are sometimes coupled. For example, Hughes et al.\ (1993)
noted that there is a trade off between $T_{\rm d}$ and the critical
frequency at which the dust becomes optically thick --- models with
dust that becomes optically thick in the far-IR can support higher
temperatures than models with dust which remains optically thin into
the mid- or near-IR regimes. This dichotomy is difficult to resolve,
even with an SED which is sampled on both the Rayleigh-Jeans and Wien
portions of the curve.
 
Here, we adopt an average value of $0.15$\,m$^2$\,kg$^{-1}$ for
$\kappa_{\rm d}$ at 800\,$\mu$m, with $\kappa_{\rm d} \propto
\nu^{\beta}$ where $\beta$, the frequency dependence of the dust grain
emissivity, is $+2.0$ (the best-fit value). It matters little whether
we adopt $\beta=+2.0$ or $\beta=+1.5$, though the latter value is
difficult to reconcile with the observed spectral index. In both cases
our mass estimates are directly comparable with those presented
elsewhere --- by Hughes et al.\ (1997) and Cimatti et al.\ (1998), for
example. To adopt $\beta=+1.5$ would raise our dust temperature
estimate by around 10\,{\sc k}, but the dust mass remains within 20
per cent of that deduced for $\beta=+2.0$.

To estimate $T_{\rm d}$, we have assumed that the dust is optically
thin.  Isothermal fits to the data then suggest that $T_{\rm d} = 40
\pm 5${\sc k} and $\beta = +2.0$ (see Figure~3).  If the true far-IR
opacity is significant, for example, if the dust becomes optically
thick at 200\,$\mu$m (as is possibly the case for Arp~220, Emerson et
al.\ 1984), then we could contrive a fit to the data which supports
$T_{\rm d} = 110$\,{\sc k} without compromising the 175- or 450-$\mu$m
data. Having adopted average dust parameters, qualified the validity
of our fits and explained the possible sources of error, a dust mass
estimate of $2 \times 10^8$\,M$_{\odot}$ then follows from equation
(2) of Hughes et al.\ (1997).
 
Although absolute measurements of dust masses are prone to large
errors, if one wants to investigate galactic evolution by comparing
the dust masses of high-redshift and low-redshift galaxies (using dust
mass as a `galactic clock', Eales \& Edmunds 1996) some of the sources
of error are removed. Uncertainties in both $H_0$ and the mass
absorption coefficient are irrelevant as long as the same values are
used at low and high redshift. Our dust mass estimate for 8C\,1435+635
is much larger than the highest dust mass derived for local spirals by
Eales \& Edmunds (1996): a factor of 3 if $q_0 = 0.5$ and a factor of
10 if $q_0 = 0$ (local ellipticals have, of course, even lower dust
masses). This immediately shows that 8C\,1435+635 is nothing like the
galaxies in the local Universe.

Table~2 lists the CO luminosities and dust masses of 8C\,1435+635 and
the other high-redshift systems that have been observed to date in CO
line and sub-mm continuum. The ratios of CO luminosity to dust mass
are quite similar to those of low-redshift galaxies. In the literature
this is often claimed as evidence that the gas/dust ratio is the same
at high redshifts as at low redshifts. This is not necessarily
correct. The metallicity of a galaxy will be a strong function of
cosmic time (Edmunds 1990), and so the hydrogen/CO ratio and the
hydrogen/dust ratio will undoubtedly change with time (a possibility
which we consider further when estimating the gas mass in \S3.3).  The
constancy of the gas/dust ratio actually tells us that the ratio of
the fraction of metals going into CO to the fraction of metals being
bound up in dust is not changing with cosmic time. Although this may
not seem a strong statement, it is reassuring that the physics of the
interstellar medium (at least in this respect) does not appear to be
changing over a large fraction of the age of the Universe.

\subsection{Instantaneous star-formation rate}

The far-IR luminosity, $L_{\rm FIR}$, suggested by dust with the
properties described in \S3.1 is $1.0 \times
10^{13}$\,L$_{\odot}$. For dust that becomes optically thick in the
far-IR, $L_{\rm FIR}$ rises to $2.7 \times 10^{13}$\,L$_{\odot}$.
 
If we assume that the energy re-emitted by the dust is initially
supplied by massive stars rather than by the AGN (the naivity of this
argument is discussed in the following section), $L_{\rm FIR}$ can be
used directly to infer the instantaneous SFR.  The far-IR luminosity
implied by our data suggest a SFR of $2000 -
5400$\,M$_{\odot}$\,yr$^{-1}$, the kind of spectacular burst that
could, if sustained, produce $10^{12}$\,M$_{\odot}$ of stars in
$<0.5$\,Gyr. This assumes a Salpeter initial mass function spanning
$1.6$ to $100$\,M$_{\odot}$, i.e.\ limited to O, B and A stars.
Reducing the lower limit to $0.1$\,M$_{\odot}$ {\em increases} the
implied SFR by a factor 3 (Thronson \& Telesco 1986).
 
In the absence of AGN heating, the ratio of $L_{\rm FIR}$ to $L_{\rm
CO}'$ provides some idea of the efficiency with which the available
molecular gas is converted into stars. For 8C\,1435+635, we find a
value $>200$\,L$_{\odot}$ ({\sc k}\,km\,s$^{-1}$\,pc$^2$)$^{-1}$,
higher than that for ultraluminous {\em IRAS} galaxies ($\sim 85$);
indeed, only quasars such as BR\,1202$-$0725 and H\,1413+117
($450-900$) and the radio galaxy, 4C\,41.17, ($>400$) rival the
efficiency with which 8C\,1435+635 apparently converts its gas into
stars. Without exception these systems host active nuclei, which
may well indicate that what we are viewing is related to AGN activity
rather than the birth of their stellar populations.

\subsection{The gas mass}

An estimate of SFR, as derived above from the far-IR luminosity, is
inevitably of dubious value in an active object such as 8C\,1435+635
simply because it is hard to distinguish whether the dust is heated
primarily by the AGN or by young massive stars. However, the issue of
the heating source is itself arguably of little importance because
even if the dust heating could be reliably attributed to young massive
stars, an estimate of the instantaneous SFR (however reliable) does
not allow one to distinguish between a violent, short-lived starburst
and a more sustained burst of star formation in which a significant
fraction of the galaxy's eventual stellar mass might be converted into
stars.
 
In practice, therefore, the most useful indicator of the evolutionary
status of 8C\,1435+635 is an estimate of the mass of gas which, at the
epoch of observation, has yet to be processed into stars.  A massive
reservoir of gas would suggest that the galaxy is extremely young
(present-day ellipticals have a large ratio of stellar mass to
molecular gas mass, where the latter is typically $<
10^{8}$\,M$_{\odot}$, e.g.\ Wiklind, Combes \& Henkel 1995, Lees et
al.\ 1991). With the information gleaned from our continuum and
spectral-line measurements, both of which are the deepest to date, we
should be able to discuss the gas mass with more confidence than is
usually the case.

Nevertheless, extrapolating from a dust mass to the total mass of
baryons in a galaxy which are {\it not} already locked up in stars at
the epoch of observation generally involves three, often rather
uncertain, steps. First a value has to be measured or adopted for the
ratio of molecular CO gas to dust. Second, a value has to be assumed
for the ratio of molecular hydrogen to CO (expected to be a function
of galaxy metallicity and hence age). Third, to the total mass of
H$_2$ must be added the mass of atomic hydrogen through the adoption
of a ratio of H\,{\sc i}/H$_2$.  To quantify the combined effect of
uncertainties in these three steps we have chosen to calculate what
can safely be regarded as rather robust lower and upper limits to the
total baryonic gas mass of 8C\,1435+635, by first adopting the highest
reasonable values for {\it all three} relevant ratios, and then adopting the
lowest reasonable values.
 
\subsubsection{Upper limit on total baryonic gas mass}
 
The highest reasonable value for the ratio of CO to dust in
8C\,1435+635 is constrained by our own failure to detect CO
emission. Using the normal Galactic calibration of $L_{\rm
CO}'$/$M({\rm H}_2)$, our non-detection corresponds to a molecular
gas/dust ratio of $M_{{\rm H}_2}/M_{\rm d} < 950$. This limit seems
entirely consistent with observations of other high-redshift galaxies
in which CO has actually been detected; observations of the lensed
quasars BR\,1202$-$0725, FSC\,10214+4724 and H\,1413+117 (Ohta et al.\
1996; Omont et al.\ 1996; Scoville et al.\ 1995; Barvainis et al.\
1997), and the $z=2.39$ radio galaxy, 53W002 (Scoville et al.\ 1997;
Hughes et al.\ 1997) yield values in the range $M_{{\rm H}_2}/M_{\rm
d} = 170-600$, after correcting to our adopted cosmology (see
Table~2).

We must then ask how reasonable it is to apply the Galactic
calibration to a galaxy at $z > 4$, particularly since metallicity is
expected to increase with age in most reasonable models of the
chemical evolution of galaxies. It is at present impossible to test
this directly, but we can gain some confidence that the ratio of
$M_{{\rm H}_2}/M_{\rm d}$ in 8C\,1435+635 is indeed very unlikely to
significantly exceed 1000 from observations of Lyman $\alpha$
absorbers.  Studies of damped Lyman $\alpha$ absorption systems
suggest $M_{{\rm H}_2}/M_{\rm d} = 400-2000$ (Fall, Pei \& McMahon
1989; Wolfe 1993) and we can be confident that 8C\,1435+635 is a more
highly evolved system than damped Lyman absorbers, the supposed
progenitors of disk galaxies, since 8C\,1435+635 is already more
luminous than an $L^*$ galaxy in starlight. We therefore adopt
$M_{{\rm H}_2}/M_{\rm d} = 1000$, and hence $M_{{\rm H}_2} = 2 \times
10^{11}$\,M$_{\odot}$ as an upper limit to the molecular gas mass of
8C\,1435+635. We thus simply sidestep the issue of whether this gas
mass arises from a CO luminosity close to our limit coupled with a
near-Galactic calibration, or whether the CO luminosity is much lower
and a low-metallicity calibration might apply.

Lastly to produce an upper limit on total baryonic gas mass we must
adopt a generous value for the ratio $M_{\rm H\,I}/M_{{\rm
H}_2}$. Barvainis et al.\ (1997) found $M_{\rm H\,I}/M_{{\rm H}_2}
\sim 4$ for H\,1413+117, although this may have been affected by
differential lensing; Andreani, Casoli \& Mirabel (1995) found $M_{\rm
H\,I}/M_{{\rm H}_2} \sim 2$ for {\em IRAS} galaxies; Wiklind et al.\
(1995) found $M_{\rm H\,I}/M_{{\rm H}_2} \sim 5$ for far-IR-selected
elliptical galaxies, whereas Lees et al.\ (1991) found $M_{\rm
H\,I}/M_{{\rm H}_2} = 1.0 \pm 0.9$). For our present purpose we
therefore adopt the largest of these values ($M_{\rm H\,I}/M_{{\rm
H}_2} \sim 5$) and hence arrive at a robust upper limit of $M_{\rm g}
= 1.2 \times 10^{12} {\rm M_{\odot}}$ for 8C\,1435+635 within our
adopted cosmology (the effect of varying cosmology is discussed in
\S4).
 
\subsubsection{Lower limit on total baryonic gas mass}
 
We now proceed, in an analogous way, to calculate a highly
conservative value for the total gas mass in 8C\,1435+635.  Models of
galactic evolution suggest that metallicity should increase with time,
and that the gas/dust ratio should decrease accordingly (Edmunds
1990).  If we take the gas/dust ratio appropriate for present-day
galaxies, we should therefore obtain a firm lower limit for the true gas
content of 8C\,1435+635.

Estimates of the gas/dust ratio in the local Universe range from
$100-150$ for the Milky Way to $\sim 1080$ for some nearby galaxies
(usually where only the warm dust has been sampled --- Devereux \&
Young 1990), and so we adopt $M_{{\rm H}_2}/M_{\rm d} = 100$
(accepting the Galactic calibration for $L_{\rm CO}'$/$M({\rm H}_2)$)
to deduce $M_{{\rm H}_2} = 2 \times 10^{10} {\rm M_{\odot}}$ as an
firm lower limit to the molecular gas mass of 8C\,1435+635.
 
Finally, we apply the adopt the lowest of the reported values for
$M_{\rm H\,I}/M_{{\rm H}_2}$ discussed above (i.e., $M_{\rm
H\,I}/M_{{\rm H}_2} = 1$) to arrive at a robust lower limit of
$M_{gas} = 4 \times 10^{10} {\rm M_{\odot}}$ for 8C\,1435+635 within
our adopted cosmology.
 
\section{Discussion: 8C\,1435+635 and 4C\,41.17 --- primeval ellipticals 
or violent mergers?}
 
What then can we conclude from our basic result that 8C\,1435+635
contains between $4 \times 10^{10}$ and $1.2 \times
10^{12}$\,M$_{\odot}$ of material which has yet to be turned into
stars at the epoch of observation corresponding to $z \simeq 4$? One
important, and {\em model-independent} point, is that our new
observations have largely served to confirm the similarity between the
sub-mm properties of 8C\,1435+635 and 4C\,41.17, despite differences
between their optical morphologies, surface brightnesses, colors and
spectra.  The fact that these two objects, both at $z \simeq 4$ and
both with comparably extreme radio luminosities, should both appear to
contain a few $10^{11}$\,M$_{\odot}$ of gas must be telling us
something rather basic.

In simple terms, the extreme far-IR luminosities, dust and gas masses
of these two sources must either be due to their extreme redshift, or
associated in some way with their extreme radio luminosities. 
Observations of larger samples of sources spanning a range in redshift and 
in radio luminosity will be required to answer this question, but it is 
interesting to consider briefly the implications of either option.

If these extreme sub-mm properties are due primarily to redshift, this
would imply that all massive ellipticals at $z \simeq 4$ could still
be in the process of forming a significant fraction of their eventual
stellar populations. Indeed, if it is assumed that the present-day
counterpart of 8C\,1435+635 is a very massive elliptical galaxy with a
stellar mass of $\sim 10^{12}$\,M$_{\odot}$, the upper end of our
derived range of gas masses is consistent with a picture in which, at
the epoch of observation, 8C\,1435+635 has yet to form the {\it vast
majority} of its eventual stellar population. This possibility gains
further credence from the fact that our derived dust mass has arguably
been minimized to a certain extent by the assumption of an Einstein-de
Sitter Universe.  In a low-density Universe the derived masses
increase by up to a factor of four, in which case even our lower limit
on the gas mass rises to $2 \times 10^{11} {\rm M_{\odot}}$ making it
difficult to argue against the conclusion that 8C\,1435+635 is rather
young. Would such a conclusion be consistent with the recent discovery
that the dominant stellar population in at least some (and arguably
all) radio galaxies is already $> 3$\,Gyr old by $z \simeq 1.5$
(Dunlop et al.\ 1996; Spinrad et al.\ 1997)? The answer is yes, and
indeed there is an interesting (if somewhat frustrating) degeneracy at
work here: reducing $q_0$ to the point where it is hard to escape the
conclusion that 8C\,1435+635 and 4C\,41.17 contain an entire galaxy's
worth of gas also stretches the cosmological timescale, allowing
3\,Gyr to elapse between $z = 4$ and $1.5$, in which case the epoch of
spectacular sub-mm emission from young elliptical galaxies would
indeed be expected to correspond to $z \ge 4$.

If our derived lower limit on the gas mass of $4 \times
10^{10}$\,M$_{\odot}$ is in fact closer to the truth then we would
seem to be forced back towards the conclusion that we are witnessing
either (i) the tail end of the formation process of massive
ellipticals, or (ii) a new injection of gas/dust from an interaction
which we see heated by the AGN or by a violent interaction-induced
starburst in what could be an otherwise well-evolved underlying galaxy
whose stars were formed at still higher redshift. This latter option
seems completely plausible because while the ages of radio sources and
their host galaxies appear to be completely decoupled (differing by
two orders of magnitude at the present day), the fact that global
radio-source activity seems to trace the cosmic star-formation history
of the Universe (Dunlop 1997) suggests that both radio sources and
starbursts are fueled in a fundamentally similar way, perhaps through
galaxy-galaxy interactions.  In this context it would perhaps not be
too surprising if the massive fuel supplies required to power the
ultra-luminous radio sources 8C\,1435+635 and 4C\,41.17 also
inevitably resulted in an associated massive burst of star-formation
activity.  Of course in a heirarchical picture of galaxy formation the
distinction between a massive merger and the final stages of galaxy
formation might be viewed as rather artificial.
 
\section{Conclusion}

While uncertainties beyond the scope of this work (e.g.\ the value of
$q_0$ and the metallicity of high redshift radio galaxies) mean that
the precise interpretation of the large mass of dust residing in
8C\,1435+635 remains ambiguous, we have demonstrated the power of
SCUBA to constrain the properties of dust in galaxies at $z > 4$.  The
radio galaxies 8C\,1435+635 and 4C\,41.17 are undeniably extreme
objects, but their very similar dust and gas masses mean that they can
certainly serve as useful benchmarks for future sub-mm studies of
high-redshift galaxies. Thus, whilst at present we cannot determine
whether the spectacular far-IR properties of 8C\,1435+635 and
4C\,41.17 are primarily due to their extreme redshifts, or their
extreme radio luminosities, the quality of our SCUBA data provides
encouragement that sub-mm observations of complete samples of radio
galaxies, spanning a range of redshifts and radio luminosities, should
be able to settle this issue in the near future. Perhaps most exciting
of all, our observations of 8C\,1435+635 serve to re-emphasize that if
most massive ellipticals do indeed form in comparably spectacular
starbursts, such objects will be easily detected by the first sub-mm
surveys of blank fields, even out to $z \simeq 10$.

\end{document}